\documentstyle[epsfig,floats,amssymb,amsbsy,aps,prl,twocolumn,tighten]{revtex}
\begin{document}
\draft 
{{}} \wideabs{
\title{Spatially selective Bragg scattering: a signature for vortices in Bose-Einstein condensates.}
\author{P. B. Blakie$^1$ and R. J. Ballagh$^{1,2}$}
\address{$^1$Department~of~Physics, University~of~Otago, P.~O.~Box~56, Dunedin, New~Zealand \\ 
$^2$ Institute for Theoretical Physics, University of Innsbruck, A-6020 Innsbruck, Austria}
\date{\today}
\maketitle 
\begin{abstract}We demonstrate that Bragg scattering from a
condensate can be sensitive to the spatial phase distribution of the initial
state. This allows preferential scattering from a selected spatial region,
and provides a robust signature for a vortex state. We develop an analytic
model which accurately describes this phenomenon and we give quantitative predictions  for
current experimental conditions. \end{abstract}

{{}} }

Bragg scattering of matter waves from laser beams is
a well known phenomenon in atom optics [e.g. \cite{martin88,berman97}] and
has proved to be an effective tool for manipulating and analysing Bose
condensates. In that context it has been used in experiments to coherently
split a condensate \cite{kozuma99,deng99}, and to
perform momentum and phonon spectroscopy \cite{MITBragg1,MITBragg2}. 
Theoretical calculations show it may be used  to extract
quasiparticle amplitudes from weakly excited condensates \cite{Brunello2000}. 
Two groups have constructed Mach-Zehnder type interferometers 
\cite{Torii2000} using sequences of Bragg pulses to probe condensate 
phase \cite{Kozuma1999,Simsarian2000,Denchlag2000}.
In all of those cases the Bragg scattering process is most easily understood in
terms of a momentum space process in which the Bragg fields transfer momentum 
$\hbar \mathbf{q}$ to each atom. In a beam-splitter, half the initial
amplitude is translated in momentum space by  $\hbar\mathbf{q}$, while in
momentum spectroscopy a selected narrow group of momentum components is
translated by $\hbar \mathbf{q}$. The spatial structure of the initial
condensate plays no particular role in the Bragg process for those cases. 

\begin{figure}[!htb]
\begin{center}
\epsfbox{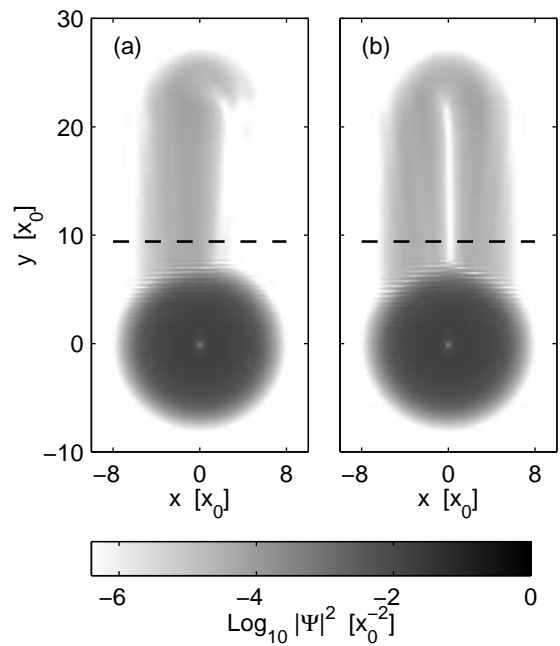}
\end{center}
\caption{Spatial density of a 2D vortex state 
at time  $t=0.6 t_0$ after excitation by a Bragg pulse with frequency detuning 
(a) $\omega=216\omega _{T}$ and (b)  $\omega =203\omega _{T}$ 
($\omega _{T} $ is the harmonic trapping frequency). Other parameters
are $V=0.2\protect\omega _{T}$,  $q=14/x_{0}$ and $ w=500w_{0} $.  
Units: time $t_{0}=1/\omega _{T}$; distance $x_{0}=\sqrt{\hbar /2m\omega _{T}} 
$; collisional interaction strength $w_{0}=\hbar \omega _{T}x_{0}^{3} $. 
The dashed line denotes a suitable position for measuring the output beam 
profile (see text).}
\label{FIG_SPAT_LOW_INTENS_SCATT}
\end{figure}

A major interest in condensates however, is that their macroscopic wavefunctions 
may be engineered into specific spatial structures. For example, the vortex state, 
which has been recently observed by two groups \cite{Matthews1999b,Madison2000a}, 
has a phase circulation of $2\pi $ about a vortex core. The most direct
means for observing such a phase distribution is by interference with a
separate well characterised matter field [e.g.
\cite{BoldaPRLDec1998,Matthews1999b}], but this may not always be
experimentally convenient. In this paper we show that under appropriate
conditions Bragg scattering is sensitive to the spatial phase distribution
of the initial condensate and therefore allows preferential scattering from
a selected spatial region. In the case of a vortex, this gives a distinctive
signature which we illustrate in Fig. \ref{FIG_SPAT_LOW_INTENS_SCATT}. There
the density distribution of a trapped condensate vortex state is shown following
application of Bragg pulse chosen according to criteria developed later in
this paper.
In Fig. \ref{FIG_SPAT_LOW_INTENS_SCATT}(a) the streaming output emerges from only
one side of the initial condensate, and gives an asymmetric density pattern
which should be detectable with current experimental technology. In Fig. 
\ref{FIG_SPAT_LOW_INTENS_SCATT}(b) where the frequency difference of the
laser fields has been changed, the streaming output field has a density node
that is an order of magnitude wider than the vortex core itself (i.e. the
healing length in the centre of the original condensate). In the following
we develop a treatment of such \textit{spatially selective Bragg scattering} 
which explains the behaviour in terms of the underlying spatial phase
sensitive mechanism and we give analytic solutions appropriate to 3D condensates.
We treat the process of Bragg scattering from a condensate
using a previously developed formalism \cite{Blakie2000}, in which the
condensate mean-field $\Psi $ (in an interaction picture) evolves in a far
detuned light-field grating formed by two plane-wave laser beams, according
to the equation 
\begin{eqnarray}i\hbar {\frac{\partial }{\partial t}}\Psi 
&=&\left[ -{\frac{\hbar ^{2}}{2m}}\nabla ^{2}+V_{T}({\bf r})+w|\Psi|^{2}\right] 
\Psi  \nonumber \\&+&\hbar V(t)\cos ({\bf q}\cdot {\bf r}-\omega t)\Psi . 
\label{EQN_GPE}
\end{eqnarray}
Here $V_{T}(\mathbf{r})$ is the trap
potential (assumed to be harmonic in the examples of this paper) 
and $w$ $($ $=4N\pi \hbar^{2}a/m)$ is the usual nonlinearity
parameter of the Gross-Pitaevskii equation. The final term in
Eq. (\ref{EQN_GPE}) describes the interaction with the Bragg grating: $\omega
$ and $\hbar \mathbf{q}$ are the two-photon detuning and recoil momentum
respectively and $V(t)=|\Omega (t)|^{2}/2\Delta  $ is twice the ac-stark
shift at the intensity peaks. For the scattering to be in the Bragg regime
the pulse length $T_{p}$ of the lasers must be long enough that the
Bragg resonance is resolved (i.e. $T_{p}\gg \omega _{q}^{-1}$, where
$\omega_{q}=\hbar q^{2}/2m$ is the recoil frequency). In order to
obtain \textit{spatially selective Bragg scattering} it is also necessary
that the average value of $V$ is sufficiently small that the total amount of
condensate scattered in time $T_{p}$ is small compared to the unscattered
amount. In addition we assume the Bragg pulse length (and the time of
observation) is shorter than a quarter trap period, to avoid the trap forces
significantly altering the momentum of the scattered beam. 

Equation (\ref{EQN_GPE}) can be directly numerically simulated, as 
for example we have shown in the two dimensional case of Fig.
\ref{FIG_SPAT_LOW_INTENS_SCATT}. We have also obtained an analytic
solution, which provides detailed insight into the underlying physics and
enables us to provide quantitative calculations over a wide range of
parameters in the three dimensional case. For the analytic solution, we
assume that the recoil momentum $\hbar \mathbf{q}$ of the Bragg grating is
much larger than the momentum width of the initial condensate (which is
centred about zero momentum), and also assume that the
detuning $\omega $ is  close to the Bragg resonance. This means
the scattered wavepacket is well separated in momentum space from the
initial state and a slowly varying envelope approximation can be made \cite{Trippenbach2000}
so that we write
 
\begin{equation}\Psi ({\bf r},t)=\psi _{0}({\bf r},t)+\psi
_{1}({\bf r},t)\,e^{i({\bf q}\cdot {\bf r}-\omega t)}. 
\label{EQ_SVEA}
\end{equation}
In Eq. (\ref{EQ_SVEA})
$\psi _{1}$ is the scattered wavepacket while $\psi _{0}$, 
which we call the mother condensate, represents a condensate with a 
momentum wavepacket centered on zero. At $t=0$, $\psi_0$ is exactly the 
initial condensate. Substituting Eq. (\ref{EQ_SVEA}) into Eq. 
(\ref{EQN_GPE}) and projecting into orthogonal regions of momentum
space, we obtain to first order in $\psi _{1}$ the coupled equations
\begin{eqnarray}i\hbar {\frac{\partial }{\partial t}}\psi
_{0} &=&\Big[ -{\frac{\hbar ^{2}}{2m}}\nabla ^{2}+V_{T}+w|\psi
_{0}|^{2}\Big] \psi _{0}+{\frac{\hbar V(t)}{2}}\psi _{1}, 
\label{EQN_Full2State1} \\
i\hbar {\frac{\partial }{\partial t}}\psi _{1}
&=&\Big[ -{\frac{\hbar ^{2}}{2m}}\nabla ^{2} -\hbar\delta-i\hbar {\bf Q}\cdot \boldsymbol{\nabla}
+V_{T}  \label{EQN_Full2State2} \\
 &+& 2w|\psi_{0}|^{2}\Big] \psi_1   
+{\frac{\hbar V(t)}{2}}\psi_0, \nonumber 
\end{eqnarray}
where ${\bf Q}=\hbar
{\bf q}/m$ is the velocity of the scattered atoms, $\delta =\omega -\omega
_{q}$ and we have neglected the rapidly oscillating term $\psi _{0}^{2}\psi
_{1}^{\ast }$. Our interest is in the scattered state $\psi_{1}$ which is
assumed to be small, so we can neglect the scattering from $\psi _{1}$ back
to $\psi _{0}$ (i.e. $V(t)\psi _{1}$ in Eq. (\ref{EQN_Full2State1})). This
means that  $\psi _{0}$ then evolves according to the usual
Gross-Pitaevskii equation, and typically we choose $\psi_{0}$ to be an
eigenstate of that equation.  We can also ignore the $ \nabla^{2}$ term
in Eq. (\ref{EQN_Full2State2}) which describes momentum diffusion about the
centre of momentum ${\bf q}$ of the $\psi _{1}$ packet, and is small on
the time scales we consider.  Eq. (\ref{EQN_Full2State2}) can now be
solved to give 
\begin{equation}\psi _{1}({\bf r},t)=-{\frac{i}{2}}\int ds\,
e^{-iK({\bf r},t,s)}V(s)\psi _{0}\left( {\bf r}\!+\!{\bf Q}(s\!-\!t),s\right),
\label{EQN_Approx_soln}
\end{equation}
where
\begin{eqnarray}K({\bf r},t,s)&=& (s-t)\delta +{\frac{1}{\hbar
}}\int_{s}^{t}ds^{\prime }\,\Big[ V_{T}({\bf r}+{\bf Q}(s^{\prime
}-t))\nonumber\\
&&+2w|\psi _{0}({\bf r}+{\bf Q}(s^{\prime }-t),s^{\prime })|^{2}\Big]\,.
\label{EQN_Aprrox_soln_integrating_factor}
\end{eqnarray}
Eq. (\ref{EQN_Approx_soln}) allows us to visualise the formation of
the scattered state as follows. As the scattered packet moves across the
mother condensate, the amplitude $\psi _{1}$ at a given point (stationary in
the frame moving with velocity ${\bf Q}$ ) is built up from the sum
of contributions coupled in from successive points along the mother
condensate. The contribution coupled in at time $s$ from a position
${\bf r}+{\bf Q}(s-t)$ on the mother condensate evolves to  time $t$ with
the propagator $\exp (-iK({\bf r},t,s)).$ The scattered state at ${\bf r}$
and $t$ will be appreciable only if the contributions constructively
interfere. Writing $\psi _{0}({\bf r,}s{\bf )}$ in terms of amplitude and
phase $A_{0}({\bf r,}s)\exp (iS_{0}({\bf r},s))$, then the condition for
an appreciable scattered state to form is that the phase $\Theta
({\mathbf r},t,s)=-K({\bf r},t,s)+S_{0}({\bf r}+{\bf Q}(s^{\prime}-t))$ be stationary. We take 
$\psi_0$ to be a stationary eigenstate, i.e  $\psi
_{0}({\bf r},s)=A_{0}({\bf r})\exp (i(S_{0}({\bf r})-\mu _{0}s)),$ where
$\mu _{0}$ is the eigenvalue of the mother condensate. Furthermore we make use of the 
Thomas Fermi solution $w[A_0]^2=\mu_0+V_T$ so then
\begin{equation}
K\approx(t-s)(\mu_0\!-\!\delta)+\frac{w}{\hbar }\int_{s}^{t}
ds^{\prime }\left[A_{0}({\bf r}+{\bf Q}(s^{\prime}\!-\!t))\right]^{2} \label{EQN_TFTheta}.
\end{equation}
The condition for
stationary phase $d\Theta ({\mathbf r},t,s)/ds=0$ then gives the \textit{generalised
Bragg resonance condition }
\begin{equation}\delta \approx \left[
\frac{w}{\hbar }\left[A_{0}({\bf R})\right]^{2}+\nabla _{{\bf R}}S_0({\bf R})\cdot {\bf Q}\,
\right]_{{\bf R}={\bf r}+{\bf Q}(s-t)}. 
\label{EQN_BraggReson}
\end{equation}
If $\Theta $ has sufficiently
large spatial curvature, one particular time $s $ may dominate the
stationary phase contribution. However for the cases of a ground or 
vortex initial state, $\Theta $ varies sufficiently slowly
that most contributions from along the line ${\bf R}={\bf r}+{\bf Q}(s-t)$
on the initial condensate can be in phase. For an initial ground state, Eq.
(\ref{EQN_BraggReson}) reduces in the linear case to the usual Bragg
resonance condition $\delta =0$ , while for the nonlinear case, $\delta
=\left\{ w|\psi _{0}({\bf R})|^{2}/\hbar \right\} $ in which the braces
indicate that the precise value of the shift is obtained by a suitable
average along the path ${\bf R}$ \cite{Note}. For a general initial state the second term in Eq.
(\ref{EQN_BraggReson}), which can be interpreted as a Doppler condition,
must be included. 

The full spatial solution for the scattered state 
contains a great deal of information, but we can extract a useful and compact
signature by considering the steady-state density profile of the scattered
beam at the edge of the initial condensate. In the 2D
simulation of Fig. \ref{FIG_SPAT_LOW_INTENS_SCATT} for example, the
steady-state density profile could be measured along the dashed line. For
definiteness, we shall take the case where ${\bf q}$ is in the $y$
direction. We define the steady-state density profile of the scattered state to be
\begin{equation}{\mathcal D}_{q}(x,z,\omega )=|\psi_1(x,y=R,z,\,\tau_{S})|^{2},
\label{EQN:def_BeamProf}
\end{equation}
where the distance $R$ is  sufficiently large that the density of the 
mother condensate in the plane $(x,y=R,z)$ is negligible 
(i.e ${\mathcal D}_{q}\approx|\Psi|^2$). For strongly interacting condensates 
it is suitable to use $R=R_{TF}$, where $R_{TF}$ is the Thomas-Fermi radius 
(in the $x$$y$-plane). $\tau _{S}$, the time to reach steady state can be estimated as
 $2R/Q$, i.e the time for the scattered state to
traverse the mother condensate.  Since experiments
would usually measure the density projected along the $z$ axis (the line of
viewing), we define the projected \textit{steady-state density profile}
\begin{equation}
 D_{q}(x,\omega )=\int dz\,{\mathcal D}_{q}(x,z,\omega ), 
\end{equation}
 to characterise the results of the Bragg scattering. 

We begin by considering the case of two spatial dimensions, 
which contains the essence
of the physics. 
\begin{figure}[!htb] 
\begin{center}
\epsfbox{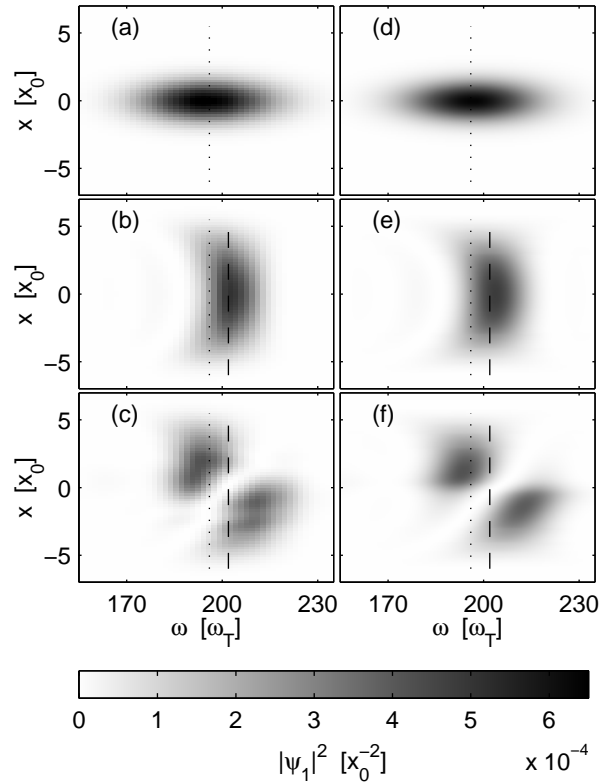}
\end{center}
\caption{Steady-state Bragg scattered density profiles in 2D for (a) noninteracting
ground state \protect\cite{Noteni}, (b) condensate ground state ($w=500w_0$), (c)  condensate vortex
state ($w=500w_0$, $m=-1$). Measurements are made at $t=0.6t_0$ for a
Bragg field with $V = 1\omega_T$, $q=14/x_0$. Frames (d)-(f) give  the analytic
solution of Eq. (\ref{EQN_Approx_soln}) for the cases  corresponding to (a)-(c)
respectively.  Dotted lines, free particle resonant
frequency  ($\omega_q=\hbar q^2/2m$);  dashed lines, 2D nonlinear
shifted frequency  ($\omega_{nl}=\omega_q+2\mu_0/3$) \protect\cite{Note}.}  
\label{FIG_2D_beam_cross_section_comparison}
\end{figure}
Fig. \ref{FIG_2D_beam_cross_section_comparison} shows $D_q(x,\omega)$ 
calculated from the full 2D numerical solution of Eq. (\ref{EQN_GPE}), for cases
where the initial state is (a) a noninteracting ground state (b) a
condensate ground state, (c) a condensate vortex state.   In Fig.
\ref{FIG_2D_beam_cross_section_comparison} (d)-(f) we provide the
comparison to the analytic solutions for the same cases, and it is apparent
that agreement is very good. We note that in the analytic solutions a 
Thomas Fermi approximation is made for $\psi_0$. For vortex 
states, the additional centrifugal potential $m^2/(x^{2}+y^{2})$ is included 
and the phase is $S_0({\mathbf r})=m\arctan(y/x)$.

The results for the noninteracting ground state (Fig.
\ref{FIG_2D_beam_cross_section_comparison}(a) or (d)) show, as expected, that 
the scattered beam is greatest (i.e. Bragg scattering is
resonant) when $\omega =\omega _{q}$ (i.e. $\delta =0$). 
The frequency width of $D_q(x,\omega )$  can
be estimated using a simple Fourier analysis on the integral in Eq.
(\ref{EQN_Approx_soln}), to be $\Delta _{D}\sim Q/R_{TF}$. 
In Fig. \ref{FIG_2D_beam_cross_section_comparison}(b) [or
(e)] the effects of the condensate nonlinearity appear. At the centre of the
scattered profile ($x=0$) the resonant frequency for Bragg scattering has
been shifted by $.84\mu $   which is close to the value of $\left\{w|\psi
_{0}({\bf R})|^{2}/\hbar \right\} $ averaged along the centre line of the
mother condensate ($=$$4\mu_0/5$). The shift at the spatial edges of the scattered beam 
($x=\pm R_{TF}$) is less, because the mother condensate has lower
average density along the appropriate lines ${\bf R}={\bf r}+{\bf
Q}(s-t)$, and so $D_q(x,\omega )$ has a crescent shape. The frequency 
width $\Delta _{D}$ (at $x=0$) is smaller than for the noninteracting ground state 
case (a), due mainly to the increased spatial width $R_{TF}$ of the mother condensate.
The most significant result for this paper is Fig. 
\ref{FIG_2D_beam_cross_section_comparison}(c), namely the vortex
signature. At the `resonant' frequency (indicated by the dashed line), the
scattered density profile is essentially spatially symmetric (see also Fig. 
\ref{FIG_SPAT_LOW_INTENS_SCATT}(b)), but at other 
frequencies the scattering is spatially asymmetric. 
This asymmetry, which we emphasize arises from the spatial phase asymmetry
of the vortex,  is robust, being present for a wide range of frequencies.
The density node at the centre of the beam ($x=0$) at the resonant
frequency also arises from the phase asymmetry: as the scattered
wavepacket passes over the vortex core, the contributions from the mother
condensate change phase sharply by $\pi $ and thus cancel.
\begin{figure}[!htb]  
\begin{center}
\epsfbox{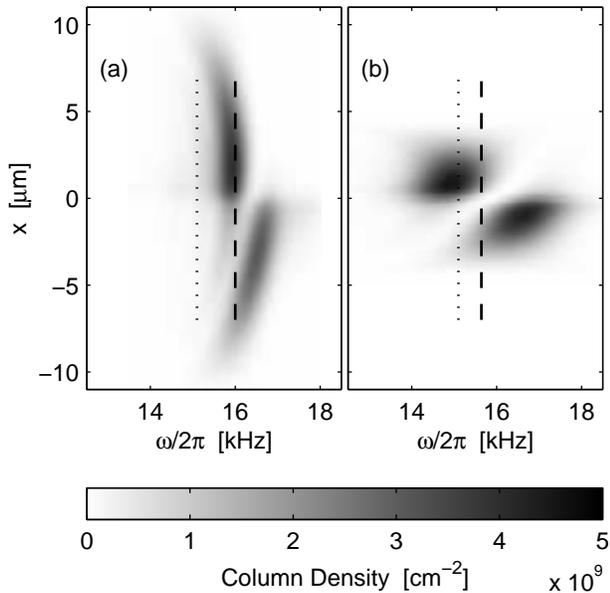}
\end{center}
\caption{Projected steady-state Bragg scattered density profiles from a condensate of $2\!\times\!10^5$
Rb$^{87}$ atoms in an $m=-1$ central vortex in (a) prolate trap with
$\omega_T=2\pi\!\times\!50Hz$, aspect ratio $\lambda=\sqrt{8}$,
$\mu_0\approx31.6\omega_T$; (b) oblate trap  with
$\omega_T=2\pi\!\times\!100Hz$, $\lambda=1/10$, $\mu_0\approx9.5\omega_T$.  Dotted
lines, free particle resonant frequency ($\omega_q=2\pi\!\times\!15kHz$);
dashed lines, 3D nonlinear shifted frequency
($\omega_{nl}=\omega_q+4\mu_0/7$). Bragg field ($V\!=\!2\pi\!\times\!50Hz$) provided by counter-propagating
lasers of approximately $780nm$. }
\label{FIG_3D_beam_cross_sections}
\end{figure}
Our results can also be extended into three dimensions by using the
analytic solution Eq. (\ref{EQN_Approx_soln}). In Fig. 
\ref{FIG_3D_beam_cross_sections} we present the
behaviour of $D_{q}(x,\omega )$ from a vortex condensate of
$2\!\times\!10^{5}$ Rubidium atoms in both oblate and prolate traps. The features
discussed in the previous paragraphs are unchanged in three 
dimensions. It is worth noting that our results can be readily 
connected to the well known dynamic structure factor 
$S({\bf q},\omega )$ which has been previously used to
characterise the results of Bragg scattering
\cite{MITBragg1,MITBragg2,Zambelli}. One can show, within perturbation
theory, that
\begin{equation}S({\bf q},\omega )\sim \int dx\,D_{q}(x,\omega)\,.
\label{EQN_relationship_to_dsf}
\end{equation}
In other words, the dynamic structure factor simply projects out all the spatial
information of the steady-state density profile.  
 
Although we have concentrated on the steady state density profile
our analytic solution contains other
interesting results. For example, with a Bragg pulse of length $T_p\ll\tau_s$,
the scattered state is itself a vortex, and a sequence of such pulses would produce a
sequence of vortices streaming out from the mother condensate, until 
it becomes too depleted.

In summary we have shown that under appropriate conditions Bragg 
scattering is sensitive to the spatial phase dependence of the 
initial matter field state, and therefore allows preferential 
scattering from a selected spatial region. We have developed an 
analytic model which accurately describes this phenomenon and 
explains the underlying mechanisms. When applied to a vortex state, 
a robust signature is obtained, for both oblate and prolate traps.

This research was supported by the Marsden Fund of New Zealand under contract PVT902. 
RJB is grateful for the hospitality of the Institute for Theoretical Physics, 
University of Innsbruck, where part of this research was carried out, and for the 
support of the Austrian Science Foundation.


\begin{thebibliography}{99}
\bibitem{Note} This shift, suitably averaged over all paths through the
condensate corresponds to the shift measured in \cite{MITBragg1,MITBragg2}. The
Thomas Fermi values for 2D and 3D are $2\mu_0/3$ and  $4\mu_0/7$ 
respectively (see \cite{MITBragg1}).
\bibitem{Noteni} For the case $w=0$ we use a gaussian wavefunction for
$\psi_0$,  with $\mu_0=0$.
\end{thebibliography}
\end{document}